# Neutron-induced opalescence in silica glass

Vasiliev S.A.[1], Kashaykin P.F.[1], Tomashuk A.L.[1]

[1]*Prokhorov General Physics Institute of the Russian Academy of Sciences, Dianov Fiber Optics Research Center, Moscow, Russia*

## 1. Abstract

Light scattering intensity distribution along the optical fiber length is analyzed, which was measured by optical time domain reflectometry (OTDR) during and after fiber irradiation in a nuclear reactor. The OTDR traces prove to contain two components associated with either radiation-induced attenuation (RIA), or with Rayleigh scattering coefficient (RSC) variation. The RSC dependence on the fast-neutron fluence is found to be strongly non-monotonic with a ~10-time increase at a fast-neutron fluence of ~$10^{18}$–$10^{19}$ $n_f/cm^2$ and a fall-off with further irradiation. Such RSC variations are interpreted in terms of thermal spikes – silica regions ~10 nm in size with higher fictive temperature and density, which arise in silica as the result of fast-neutron bombardment. In the first stage of irradiation, thermal spikes are factors of silica inhomogeneity to increase RSC; however, as soon as most silica has been occupied by thermal spikes, further irradiation only enhances silica homogeneity to reduce RSC. The RSC increase effect in the first stage of fast-neutron irradiation is argued to be similar to the anomalous scattering buildup near the critical point of the liquid-vapor transition known as critical opalescence. For his reason, the RSC increase due to fast-neutron irradiation can be named *neutron-induced opalescence* (NIO).

## 2. Introduction

Crystalline and amorphous modifications of silicon dioxide have wide practical applications in science and technology as optical materials. Their relatively high radiation resistance together with other advantages (high optical, mechanical, and thermal characteristics, abundance in nature, ease of manufacture, resistance to aggressive environments, durability, etc.) make these materials irreplaceable in many uses in nuclear and thermonuclear reactors. In recent decades, reactor applications of silica optical fibers in sensors of physical quantities as sensing elements and as transport fibers have acquired great importance [1, 2].

Reactor irradiation of silica by gamma rays and neutrons is accompanied by rearrangement of the glass network, which includes microscopic changes (occurrence of points defects [3]) and macroscopic changes (density and refractive index variation, etc. [4, 5, 6]). The processes of radiation-induced ionization and damage show up in various optical properties [7, 8], electron spin resonance [3, 9], Raman [10, 11, 12] and X-ray [13] scattering, etc.

Solid materials (e.g. metals) are known to demonstrate swelling as the dominating effect under prolonged neutron irradiation [14, 15]. However, some of them (e.g. silica) transform into a denser state (so-called 'metamict' state or phase in the case of silica). In the metamict state, silica density and refractive index are known to increase by ~3% [4, 16, 17].

Neutron impact leads to displacement of atoms in the silica glass network, with some portion of the kinetic energy dissipating to cause local heating of the glass to temperatures of

~$10^4$°C [18]. The heated glass network fragments, ~10 nm in size, quickly cool down within $10^{-11}$ sec to acquire a higher fictive temperature [5, 19]. These network fragments ('thermal spikes' [20, 21]) possess a higher density, which manifests itself in a Si-O-Si angle reduction by 10° [5] and an increase in the 3-member ring concentration [5, 10]. Microscopic manifestations of the silica rearrangement show up at fast-neutron fluence of $10^{17}$–$10^{18}$ $n_f$/cm$^2$, whereas the transition process to the metamict state reaches saturation at >$10^{20}$ $n_f$/cm$^2$ [4]. It is known that the metamict state of silica glass occurs when the energy density deposited in the form of nuclear displacement becomes as high as ~$10^{24}$ eV/cm$^3$ [22]. Such deposited energy is achieved at a fast-neutron fluence of about $10^{20}$ $n_f$/cm$^2$ [4], which approaches 1 dpa for silica [28]. This means that in the metamict state, almost all atoms of the glass network have undergone displacement as a consequence of neutron collisions.

Nuclear displacement due to fast-neutron impact transmits three orders of magnitude greater energy to silica than pure ionizing radiation [22]. For this reason, usually, only neutron fluence is considered as a measure of damage inflicted on the silica network by mixed gamma-neutron irradiation.

Optical time domain reflectometry (OTDR) is a widespread method to register inherent and induced imperfections along the optical fiber length (loss, splicing points, fiber breaks, etc.). OTDR is also used to measure the distribution of radiation-induced attenuation (RIA) along the fiber length situated in the nuclear reactor active zone. A relatively short fiber length (~1 m) is sufficient to study RIA in the reactor, but high-accuracy measurement of RIA distribution by OTDR are impossible on such a short fiber length [23, 24, 25].

In most papers on reactor irradiation, OTDR traces of both pure-silica-core fibers [26, 27, 28] and standard germanosilicate fibers [29, 30] featured hump-like distortions in the region of the fiber entry in and exit from the reactor active zone. In [27], the "humps" were supposed to be due to Fresnel light reflection at the interface between irradiated and unirradiated fiber sections, in which radiation-induced refractive index variation is different [4]. At the same time, the radiation field in the reactor active zone varies smoothly as compared to wavelength, and, therefore, the transition between the irradiated and unirradiated fiber lengths is rather long (~10 cm) and can hardly act as a Fresnel reflector. Below, we prove that it is neutron-induced Rayleigh scattering coefficient (RSC) variation that is responsible for the hump-like distortions of the OTDR traces. It should be noted that in a few earlier papers [28, 30, 31, 32], the hypothesis of the RSC variation during reactor irradiation was mentioned as a possible explanation of the evolution of the OTDR traces; however, no analysis of the shape of the OTDR traces was carried out.

In this paper, we analyze the shapes of the OTDR traces obtained in reactor-irradiated fibers in [26] and [27] after subtracting the share of the OTDR signal associated with RIA. As the result, we obtain the RIA-subtracted OTDR traces associated with RSC only. As shown below, the fluence dependence of the RIA-subtracted OTDR traces unambiguously points to RSC variation as the origin of the hump-like distortion.

3. **Results**

Radiation-induced hump-like distortion of OTDR traces was observed in many studies, of which the OTDR traces most convenient for processing were obtained in two papers [26, 27]. In [26], single-mode fibers with a refractive index difference between the pure silica core and fluorosilicate cladding of ~0.01 were irradiated in the WWR-K reactor active zone to the fast-neutron fluence of $1.8\times10^{20}$ $n_f$/cm$^2$ at a fast-neutron flux of $1.08\times10^{14}$ $n_f$/(cm$^2$s). A detailed description of the fiber and irradiation parameters can be found in [33].

Fig. 1a shows the OTDR traces measured by two-wavelength reflectometer JDSU MTS 6000 in one of the fibers two years after irradiation completion (blue and red circles). The traces between coordinates 54 and 56 m correspond to the fiber part located in the reactor active zone (filled area), where the fluence was nearly constant along the fiber. As follows from the slope of the OTDR traces, residual RIA in this fiber part amounted to ~1 and ~2 dB/m at λ=1.31 and 1.55 μm, respectively. In the ~1-meter-long transition regions (hatched areas), the fast-neutron fluence monotonically decreased; nevertheless, the OTDR traces showed up a characteristic hump-like distortion at both wavelengths.

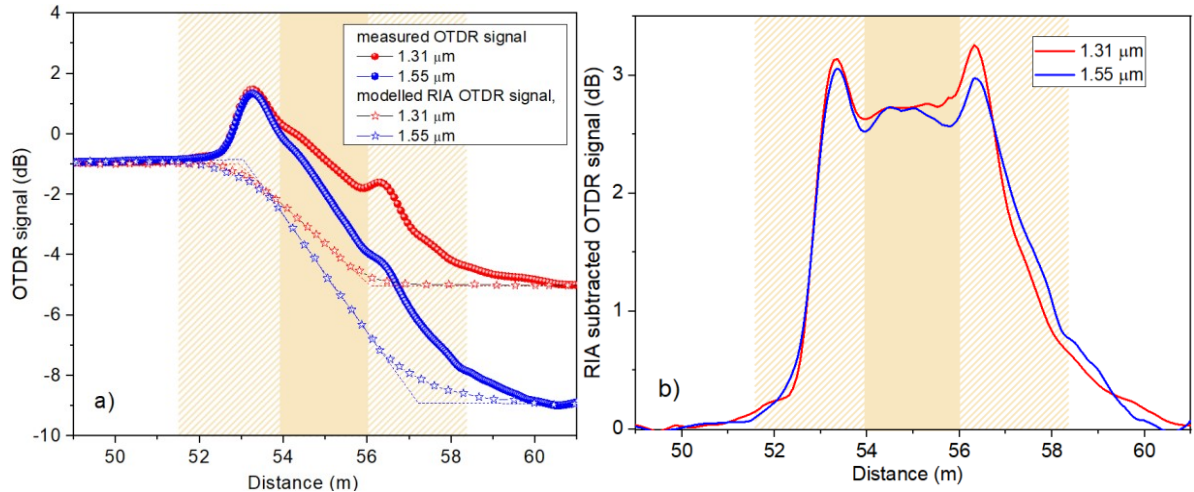

Fig. 1 a) OTDR traces measured in [26] at λ=1.31 μm (red circles) and 1.55 μm (blue circles); RIA-associated contributions to the OTDR traces (stars of the corresponding color); RIA-associated contributions in the assumption of infinite spatial resolution and abrupt fluence variation along the fiber length (dashed lines).
b) RIA-subtracted OTDR traces.
Filled areas in both figures show the reactor active zone, hatched areas, the transition regions.

We assumed the RIA contribution to be linear in the reactor active zone (constant slope of the OTDR traces) and to smoothly convert into horizontal lines (RIA=0) beyond the transition regions. To separate the Rayleigh scattering contribution to the OTDR traces, we subtracted the RIA-associated OTDR traces depicted by stars in Fig. 1a from the experimental OTDR traces, the RIA-subtracted OTDR traces obtained being shown in Fig. 1b. One can see that during the irradiation campaign, RSC increased approximately equally over the whole reactor active zone (54 – 56 m) with uniform fast-neutron fluence. The RSC increase proved to be equal at both wavelengths (~2.7 dB in the OTDR units, i.e. a factor of ~3.5 in reality). The presence of the OTDR signal maxima in the transition regions, where the fluence smoothly went down, suggests that the RSC induced by reactor irradiation varied non-monotonically with the fast neutron fluence. Note that the width of the maxima was limited by the OTDR spatial resolution of ~0.6 m; therefore, their real amplitude could be larger than that shown in Fig. 1b.

We also analyzed the OTDR traces obtained in a similar experiment on a single-mode pure-silica-core fluorosilicate-cladding fiber at a fast-neutron flux of $1.7 \times 10^{13}$ $n_f$/(cm$^2$s) [27]. Experiments in [26] and [27] differed slightly in that the fiber length situated in the uniform radiation field of the reactor active zone was a little less in the latter case (~1 m) and the irradiation temperature was 50°C as compared to 180°C in [26]. In addition, the OTDR traces in [27] were taken immediately during irradiation and at wavelength λ=0.85 μm. However, the distinctions between those experiments proved to be insignificant.

Fig. 2a reproduces the OTDR traces measured in [27] at nine time points during the reactor irradiation. As an illustration to our analysis, Fig. 2a also shows the assumed RIA-associated contribution to OTDR trace #9 (yellow line with circles) together with its shape in the virtual case of unlimited OTDR spatial resolution and abrupt fluence variation along the fiber length (dashed line). Note that the measured OTDR trace #9 (yellow curve) would coincide with the RIA-associated contribution to the OTDR trace were it not for the radiation-induced RSC variation. The reactor active zone with uniform irradiation and the transition regions, of which the boundaries were approximately deduced from the shapes of the traces, are also shown in Figs. 2 a,b by filled and hatched areas, respectively.

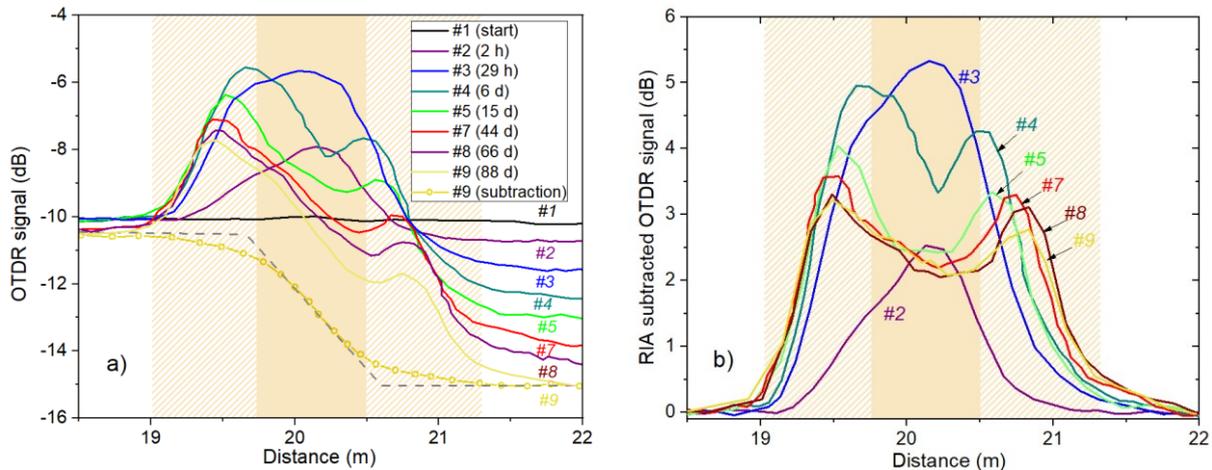

*Fig. 2 a) OTDR-traces measured in [27] during reactor irradiation, assumed RIA-associated contribution to OTDR trace #9 (line with circles), RIA-associated contributions to the OTDR trace #9 in the assumption of infinite spatial resolution and abrupt fluence variation along the fiber length (dashed line).*
*b) The same traces after subtraction of the RIA-associated contribution.*
*Filled area shows the reactor active zone, hatched areas, the transition regions.*

Fig. 2b depicts the evolution of the RIA-subtracted OTDR traces during irradiation. One can see once again that the RSC dependence on fluence is essentially non-monotonic, which explains the peaks in the transition regions for the last five traces (#4, 5, 7, 8, 9). At the same time, for traces #2 and #3, RSC proved to be lower in the periphery than at the very center of the active zone. For trace #3 (fast-neutron fluence of $\sim 2 \times 10^{18}$ $n_f/cm^2$), RSC reached maximum, which was proved to be more than 10 times (>5 dB in the OTDR units) greater than the initial RSC. As the fluence increased further (traces #5, 7, 8, 9), RSC went down at the center and tended to saturation at the level of ~ 2 dB (i.e. RSC remained ~5 times higher than the initial value).

In Fig. 2b, it is also evident that the OTDR signal maximum located at the active zone center at the beginning of irradiation split up into two hump-like peaks shifting towards the transition regions with further irradiation. If the maxima were due to Fresnel reflection, they would remain at the same coordinates throughout the irradiation campaign, their amplitudes increasing as the refractive index difference between the 'irradiated' and 'unirradiated' fiber sections increased.

Note that RSC varied with fluence in the transition region (coordinates ~19.5 and ~20.8 m, #5-#9) to a lesser extent than at the active zone center (traces #3, #4). This fact could be due to either a lower fast-neutron flux, or a greater flux gradient in the periphery. In the latter case, the limited OTDR spatial resolution could reduce the real peaks' height.

### 4. Discussion

Our analysis proves that fast-neutron bombardment changes the Rayleigh scattering intensity in pure-silica-core fibers, the RSC variation with fluence being non-monotonic.

The RSC fluence dependence is illustrated in Fig. 3 by two curves of RIA-subtracted OTDR signal variation constructed based on Figs. 1b and 2b. Black symbols correspond to the data obtained at the active zone center (Fig. 2b), whereas the red ones, to the data obtained at both active zone center and periphery (Fig. 1b). One can see that both dependences exhibited well-pronounced maxima in the region $10^{18}$ - $10^{19}$ $n_f/cm^2$ followed by saturation at ~$10^{20}$ $n_f/cm^2$.

We also constructed the angle derivative of the small-angle X-ray scattering intensity variation with fast-neutron fluence measured in bulk silica [13]. This dependence (red curve in Fig. 3) proved to be similar in shape with RSC curves, its maximum lying in the same fast-neutron fluence region. The well-known dependence of silica density variation with fast neutron fluence [4] also shown in Fig. 3 (blue curve) is of a different shape: density increases monotonically up to $5\times10^{19}$ $n_f/cm^2$ and then exhibits some fall-off with further irradiation.

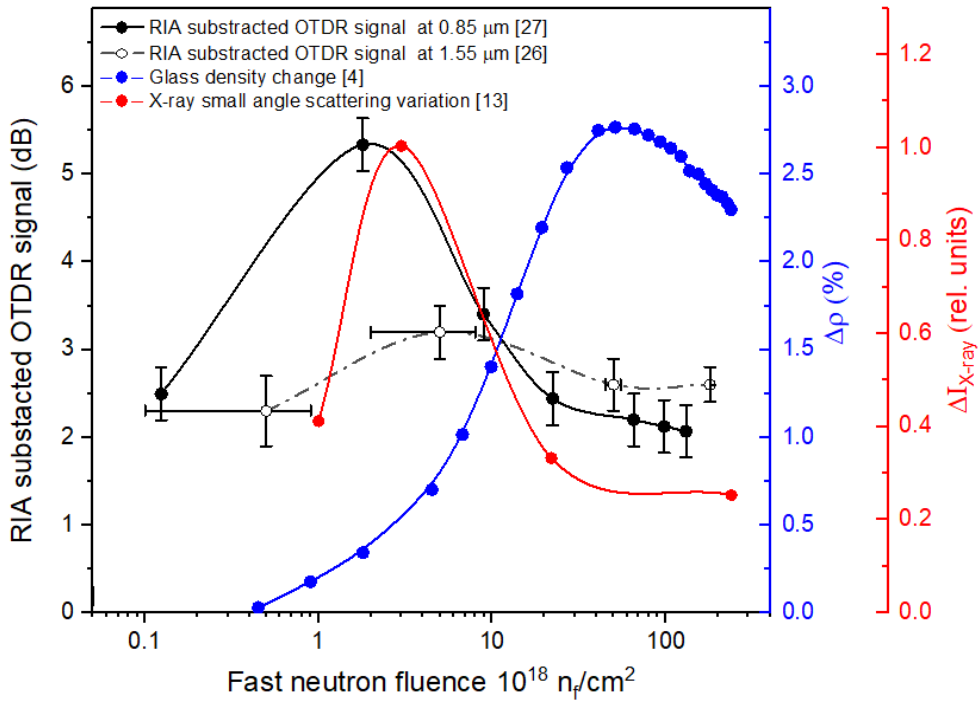

*Fig. 3. Fast neutron fluence dependences of RIA-subtracted OTDR signal variations (black symbols) [26, 27], small-angle X-ray scattering intensity variation (red curve) [13], and silica glass density (blue curve) [4].*

Structural changes in silica during irradiation seen as peaks in the region $10^{17}$ – $10^{20}$ $n_f/cm^2$ (Fig. 3) can be explained in terms of neutron-induced thermal spikes. These changes fall, apparently, into two stages: in the first stage (~$10^{17}$ - $10^{19}$ $n_f/cm^2$), the neutron-induced thermal spikes act as inhomogeneities increasing Rayleigh and small-angle X-ray scatterings. However, as soon as the glass has been occupied by spikes to a sufficient extent, further irradiation only enhances glass homogeneity by further increasing the spike concentration. As the result, the intensity of the scatterings starts to go down tending to saturation at metamict state ($>10^{20}$ $n_f$/cm$^2$). It is worth noting that the assumption of two stages of the neutron-induced silica rearrangement exhibited by small-angle X-ray scattering was made in [13].

The phenomenon of neutron-induced anomalous scattering in silica is analogous to opalescence, dramatic enhancement of light scattering, in particular, near the critical point of the liquid-vapor transition [34] , or at some other structural transitions in liquids and solids [35, 36, 37, 38]. Opalescence is due to a significant increase in density fluctuations; therefore, it would be reasonable to interpret the phenomenon of Rayleigh scattering rise revealed in this paper as a peculiar kind of opalescence, namely *neutron-induced opalescence* (NIO).

## 5. Conclusion

It was shown in this paper that the Rayleigh scattering coefficient (RSC) of silica optical fibers varies under the action of fast-neutrons and this variation is non-monotonic with fluence. We argued the RSC variation to be associated with thermal spikes arising in the fiber silica network as the result of the fast-neutron bombardment. Thermal spikes are known to be microscopic regions with a significantly enhanced fictive temperature and, therefore, a higher density; for this reason, in the first stage of fast-neutron-induced silica-network modification, spikes become a factor of network inhomogeneity to increase RSC. However, starting from a certain fluence, the situation is reversed and further irradiation producing new spikes only enhances network homogeneity to lower RSC. Thereafter, RSC continues to smoothly decrease tending to saturation in the metamict state.

The non-monotonic RSC dependence on fast-neutron fluence explains the hump-like distortion of the OTDR traces of silica optical fibers irradiated in the active zone of nuclear reactors, which was observed in a number of papers at the fiber points of entry to and exit from the reactor active zone in the fluence region of $\sim 10^{18}$ - $10^{19}$ $n_f$/cm$^2$. Similar behavior with respect to fast-neutron fluence and with a close peak position was demonstrated by the angle derivative of the small-angle X-ray scattering measured in bulk silica to suggest that it is thermal spikes that are responsible for this effect as well.

The phenomenon of neutron-induced anomalous light scattering in silica studied in this paper is analogous to opalescence, which consists in strong enhancement of light scattering in some liquids and solids under certain conditions. Based on this similarity, we propose to name the phenomenon revealed in this paper as *neutron-induced opalescence.*

**Acknowledgements**

The authors would like to thank the staff of the Large-scale research facilities "Fibers" (UNU Fibers) of GPI RAS for the fabrication of the used fiber.